# Deep Else: A Critical Framework for AI Art

Dejan Grba

Interdisciplinary Graduate Center, Digital Art Program, University of Arts, 11000 Belgrade, Serbia; dejan.grba@gmail.com

**Abstract:** From a small community of pioneering artists who experimented with artificial intelligence (AI) in the 1970s, AI art has expanded, gained visibility, and attained socio-cultural relevance since the second half of the 2010s. Its topics, methodologies, presentational formats, and implications are closely related to a range of disciplines engaged in the research and application of AI. In this paper, I present a comprehensive framework for the critical exploration of AI art. It comprises the context of AI art, its prominent poetic features, major issues, and possible directions. I address the poetic, expressive, and ethical layers of AI art practices within the context of contemporary art, AI research, and related disciplines. I focus on the works that exemplify poetic complexity and manifest the epistemic or political ambiguities indicative of a broader milieu of contemporary culture, AI science/technology, economy, and society. By comparing, acknowledging, and contextualizing both their accomplishments and shortcomings, I outline the prospective strategies to advance the field. The aim of this framework is to expand the existing critical discourse of AI art with new perspectives which can be used to examine the creative attributes of emerging practices and to assess their cultural significance and socio-political impact. It contributes to rethinking and redefining the art/science/technology critique in the age when the arts, together with science and technology, are becoming increasingly responsible for changing ecologies, shaping cultural values, and political normalization.

## 1. Introduction

Artists have been working with AI since the 1970s. AI art pioneers, such as Harold Cohen, Arthur Elsenaar and Remko Scha, David Cope, Peter Beyls, and Naoko Tosa, comprised a minuscule enclave within the computer art community, primarily due to the complexity and scarcity of AI systems throughout the 1970s and 1980s, but also because this period was one of the AI "winters", with reduced research funding and receding interest in the field [1]. AI research in the 1990s and 2000s provided more accessible tools for the artists to confront and compare human and machinic behavior. With uncanny robotic works that question the meaning of agency, creativity, and expression, artists such as Ken Feingold, Ken Rinaldo, Louis Philippe-Demers, Patrick Tresset, and others had articulated some of the contemporary AI art's topics.

Since the 2000s, artists such as Luke DuBois, Sam Lavigne, Sven König, Parag Kumar Mital, Kyle McDonald, Golan Levin, Julian Palacz, and others, have been creating generative and interactive works based on logical systems or statistical techniques which conceptually and technically overlap with, or belong to, AI technologies. They used natural language processing (NLP), pattern recognition, and computer vision (CV) algorithms to address various features of human perception reflected in AI, and to explore higher-level cognitive traits by interfacing human experiential learning with machine learning (ML) [2,3].

The increasing affordance of multilayered sub-symbolic ML architectures such as Deep Learning (DL), as well as the raising socio-political impact of AI in the second half of the 2010s, have facilitated the expansion of AI art [4] (pp. 2–3). Its production has been gaining momentum with the support of AI companies such as Google or OpenAI, and academic programs which have been facilitating art residencies, workshops, and conferences, while its exhibition has expanded from online venues to mainstream galleries and museums.

Contemporary AI art includes practices based on diverse creative approaches to, and various degrees of technical involvement with, ML [5] (p. 39). Its topics, methodologies, presentational formats, and implications are closely related with a range of disciplines engaged in AI research, development, and application. AI art is affected by epistemic uncertainties, conceptual challenges, conflicted paradigms, discursive issues, ethical, and socio-political problems of AI science and industry [6]. Similar to other new media art disciplines, AI art has had an ambivalent relationship with mainstream contemporary artworld (MCA); it is marked by selective marginalization and occasional exploitation, which entice artists to compromise some of their key poetic values in order to accommodate the MCA's conservative requirements for scarcity, commercial viability, and ownership [6] (pp. 252–254).

Its interdependence with AI technologies and socio-economic trends exposes AI art to a critical consideration within a broader cultural context. The existing literature comprises several studies of AI art and implicitly relevant works. For example, Melanie Mitchell in *Artificial Intelligence* (2019) [7], as well as Gary Marcus and Ernest Davis in *Rebooting AI* (2019) [8] provide a conceptual, technological, and socio-cultural critique of AI research. Michael Kearns and Aaron Roth in *The Ethical Algorithm* (2019) [9], and Matteo Pasquinelli in *How a Machine Learns and Fails* (2019) [10], address the ethical, socio-political, and cultural consequences of the AI's conceptual issues, technical imperfections, and biases. With *The Artist in the Machine* (2019) [11], Arthur I. Miller includes AI art in the examination of creativity that spans his other books [12,13]. In *AI Art* (2020) [14], Joanna Żylińska opens a multifaceted discussion of AI focusing on its influences on visual arts and culture. In the Artnodes journal special issue *AI, Arts & Design: Questioning Learning Machines* (2020) [4], edited by Andrés Burbano and Ruth West, the contributors address the issues of authorship and creative patiency (Galanter) [15], the creative modes of AI art practices (Forbes) [16], the public AI art (Mendelowitz) [17], the use of ML in visual arts (Caldas Vianna) [18], and the relationship between AI art and AI research (Tromble) [19]. In the *Atlas of AI* (2021) [20], Kate Crawford maps the less desirable reflections of human nature in the AI business, hidden behind marketing, media hype, and application interfaces. In *Understanding and Creating Art with AI* (2021) [21], Eva Cetinić and James She provide an overview of AI research that takes art as a subject matter, outline the practical and theoretical aspects of AI art, and anthologize the related publications. In *Tactical Entanglements* (2021) [22], Martin Zeilinger investigates the tactical and posthumanist values of AI art. In *Alpha Version, Delta Signature* (2020) [3], I explore the cognitive aspects of AI art practices; in *Brittle Opacity* (2021) [6], I address the ambiguities that AI art shares with AI-related creative disciplines; in *Immaterial Desires* (2021) [23], I focus on the AI art's entanglements and cultural integration; and in *The Creative Perspectives of AI Art* (2021) [24], I discuss the creative dynamics of contemporary AI art.

In this paper, I sketch a framework for the critical exploration of AI art. It comprises a contextual summary of AI art (this section); an overview of the prominent poetic features (Section 2); an outline of major issues (Section 3); and possible directions to tackle them and advance the field (Section 4). The scope includes AI art and AI-related art produced before the current AI "spring", contemporary AI art, and AI-derived mainstream contemporary art. I describe divergent methodological, exploratory, and expressive approaches of AI artists, and their effectiveness in addressing the phenomenological, epistemic, socio-political, and other aspects of AI science/tech and AI-influenced society. I also address the entanglements and cultural integration of AI art, and the ambiguities it shares with other AI-related creative disciplines. I focus on the contextually relevant works that exemplify poetic complexity and manifest the epistemic or political ambiguities indicative of a broader milieu of contemporary art, culture, AI science/technology, economy, and society. I listed the bibliographical references for the artworks discussed in more detail. I also provide numerous examples to be explored and compared by the reader. All cited artworks are well documented online and offline, and their essentials (artist's name, title of the work, and production year) suffice for a successful online query. By examining the artists' creative approaches towards the constitution of social reality through globalized AI-driven infrastructures of industry, commerce, communication, entertainment, and surveillance, I identify the conceptual, discursive, and ethical issues that affect the poetic outcomes, cultural status, and socio-political impact of AI art. This allows me to outline some of the creative, conceptual, tactical, and strategic prospects for the advancement of the field.

## 2. Poetics

This framework requires an inclusive view on the poetic features of prominent AI art practices, ranging from popular/mainstream to tactical and experimental. The discussion addresses the thematic, conceptual, methodological, technical, and presentational aspects of exemplar artworks that belong to disparate formal, thematic, or procedural categories.

The poetic scope of AI art derives from computer art and generative art, and is primarily informed by the various phenomenological aspects of sub-symbolic ML systems. It comprises the strategies that explore the epistemological boundaries and artifacts of ML architectures; sample the latent space of DL networks; aestheticize or spectacularize the renderings of ML data; and critique the conceptual, existential, or socio-political consequences of applied AI; a few works criticize AI art itself. The existing taxonomies or categorizations of AI art are useful [16,17] but should be considered as provisory due to the creative dynamics of the field, particularly with respect to AI research.

### 2.1. Creative Agency and Authorship

Themes such as creative agency, authorship, originality, and intellectual property are widely attractive to AI artists, popular with the media, and fascinating to the audience. The malleability of these notions was central to modernism and postmodernism, and artists have been addressing them with computational tools since the 1960s, so this recent surge of interest is probably due to a combination of the novelty of DL, its processual opacity, and its specific informational or formal effects. However, artistic exploration of this territory has been challenged by the AI's most pervasive ambiguity—anthropomorphism.

Anthropomorphism manifests in various forms. One is a tendency to assign human cognitive or behavioral features to non-human entities or phenomena, which often proves difficult to identify and sometimes has undesired consequences. In AI research, the anticipation of emergent intelligence is based upon a belief that software will attain intelligence and develop emotions given enough training data and computational power. Besides being technically dubious, this is an anthropocentric position, another version of the tenet that humanity is the sine qua non of the universe [19] (p. 5). It is complicated by the corporate AI's crowdsourcing of cheap, invisible, and underrecognized human labor for tasks such as dataset interpretation, classification, or annotation, whose outcomes affect ML training models or algorithms [10] (p. 7) [14] (pp. 119–127).

Anthropomorphism is broadly sensationalized in AI art discourse, for example by authors such as Arthur I. Miller who argues for the (intrinsic) creativity of AI systems [11,25]. His narratives often rely on anthropocentric expressions, such as "what neural network sees", and identify creativity in AI as generating new information [25] (pp. 247, 249). A converse form of anthropomorphic fallacy is to conflate the artists' creative agency with cumulative human creativity embedded in their tools (computers and software), which simultaneously deprives artists of their own inventiveness, and lifts the responsibility off of their creative acts. Shared between some researchers, many artists, and the media, it often exploits the trope of the ever "blurring line between artist [ghost] and machine" [26,27], and involves experiments which are supposed to establish "who is the artist" or "what is real/better art" by manipulating the preferential conditions of human subjects tasked with evaluating human- and machine-produced artefacts [25] (p. 248) [27,28]. Such experiments are often naïve or manipulative because they presume—and instruct the subjects—that their test material *is* art while omitting two fundamental distinctions: who considers something as an artwork, and why [29] (p. 102). They disregard that art is artificial by definition, and ignore the well-informed notions about a complex relationship between creative agency, authorship, and technology [3] (pp. 75–77) [5] (pp. 42–43, 47) [30–32].

2.1.1. The Elusive Artist

In his long-term project *AARON* (1973–2016), the pioneering AI artist Harold Cohen experimented with translating and extrapolating some components of human visual decision-making into a robotic drawing/painting system [33]. He had an ambiguous relationship with machinic creative agency and flirted with anthropomorphic rhetoric [34]. Not surprisingly, a highly popular segment of contemporary AI art belongs to the saccharine reiterations of Cohen's approach, in which artists "teach" their robots how to paint, such as Pindar Van Arman's *Painting Robots* (since 2006) [35] or Joanne Hastie's *Abstractions* (*Tech Art Paintings*) (since 2017) [36]. Driven by weekend painters' enthusiasm, these projects "serendipitously" merge technically competent execution with dilettante aesthetics, conceptual ineptness, and ignorance of art-historical context. The meaning of the word "art" collapses into banal, camera-driven visualizations, rendered and presented with amateurish self-confidence in a series of "progressively improved" ML systems. Anthropomorphism is also advocated within the art-academic domain, for example by Simon Colton's discussion of his project *The Painting Fool* (2012) that he hopes "will one day be taken seriously a creative artist in its own right." Its aim is to dramatically expand the artistic range of Cohen's *AARON* by introducing the software interface that could be trained by different human artists, able to critically appraise its own work, and (in future versions) the work of other artists [21] (pp. 8–9) [37] (pp. 5–6).

Fewer artists address the subtlety of this topical range. One of them is Adam Basanta. In his installation *All We'd Ever Need Is One Another* (2018) [38], a custom software randomizes the settings of two mutually facing flatbed scanners so that in every scanning cycle, each captures a slightly altered mix of the facing scanner's light and its own unfocused scanning light reflected off the opposite scanner's glass plate. The perceptual hashing algorithms then compare each scan to the images in a large database assembled by scraping images and image metadata from freely accessible online artwork repositories. If the comparison value between the scan and its most similar database image exceeds 83% based on the parameters such as aspect ratio, composition, shape, and color distribution, the software declares a "match", selects the scan for printing, and labels it according to the matching image metadata. When it selected and labeled one of the scans as *85.81%_match: Amel Chamandy 'Your World Without Paper', 2009*, Canadian artist Amel Chamandy initiated a legal action about the intellectual property rights against Basanta because of the reference to her photograph, although *85.81%_match . . .* is not for sale and Basanta apparently does not use it for direct commercial gains by any other means. *All We'd Ever Need Is One Another* disturbs the concepts of authorship, originality, and intellectual property by legitimately and consistently applying the functional logic of ML, while the intricacies of the lawsuit it triggered exemplify the intellectual and ethical issues of our tendency to crystalize the commercial rights of human creativity [22] (pp. 94–108).

Basanta and other exemplar artists such as Nao Tokui (discussed in Sections 2.2.2 and 2.4.2) or Anna Ridler (discussed in Sections 2.1.4, 2.4.2 and 2.4.3), approach AI both as a criticizable technology and a socio-political complex, and recognize the variable abstraction of technologically entangled authorship. They demonstrate that crucial aesthetic factors such as decision-making, assessment, and selection are human-driven and socially embedded regardless of the complexity or counter-intuitiveness of the tools we use for effectuating these factors. They remind us that our notion of art is a dynamic, evolving, bio-influenced, and socio-politically contextualized relational property which needs continuous cultivation.

2.1.2. Performative Aesthetizations

Performance artists who enjoy the sponsorship of corporate AI tend to emphasize dubious human-centered notions of creative agency through sleekly anesthetized mutations of earlier avant-garde practices. For example, Sougwen Chung's homo-robotic projects, such as *Drawing Operations Unit: Generation 2* (2017, supported by Bell Labs) [39], draw a comparison with Roman Verostko's algorist compositions from the 1980s and 1990s [40]. Whereas Verostko encapsulates his coding experiments with pure form into a discreet relationship between his pen-plotter and its material circumstances, Chung uses the theatricality of her collaboration with robots as a "spiritualizing force" to mystify the manual drawing process—which is by nature highly improvisational and technologically interactive.

Similarly, Huang Yi's robotic choreography *HUANG YI & KUKA* (since 2015, sponsored by KUKA) [41] spectacularizes the metaphors of harmonious human-machine interaction and mediates them safely to the comfortable spectators, while the referential Stelarc's performances since 1976, such as *Ping Body* (1996), emphasize the existential angst and uncertainty of shared participatory responsibilities between the artist, technology, and the audience who all have a certain degree of manipulative influence on each other [42] (pp. 185–190). Also sponsored by KUKA, Nigel John Stanford's musical performance *Automatica: Robots* vs. *Music* (2017) [43], can be viewed as an encore of Einstürzende Neubauten's concerts from the 1980s "spiced up" for tech-savvy cultural amnesiacs [44]. Rehearsed beyond the point of self-refutation, Stanford's "improvisations" stand in as formally polished but experientially attenuated echoes of Einstürzende's rugged guilty pleasures in sonic disruption.

With high production values and aesthetics palatable to the contemporary audience, these AI-driven acts largely evade the unfavorable comparisons with their precursors, and serve as marketing instruments for their corporate sponsors by promoting vague notions of robotically-enhanced consumerist lifestyle. Their persuasibility relies on our innate anthropocentrism, myopic retrospection, and susceptibility to spectacles.

2.1.3. The Uncanny Landscapes

The exploration of anthropomorphism in AI art often involves the uncanny appearance of artificial entities. Uncanniness is the occasional experience of perceiving a familiar object or event as unsettling, eerie, or taboo, and it can be triggered in close interaction with AI-driven imitations of human physique or behavioral patterns [45] (pp. 36–37).

Some artists approach it implicitly, for example by extracting human-like meaningfulness from the machinic textual conversation in Jonas Eltes' *Lost in Computation* (2017) [46] with reference to Ken Feingold's installations such as *If*, *Then*, *What If*, and *Sinking Feeling* (all 2001) [47]. In these works, NLP systems provide semantically plausible but ultimately senseless continuation of narrative episodes which allude to the flimsiness of the Turing test and serve as vocalized metaphors of our lives. They extend the experience of uncanny awkwardness into the absurdity of miscommunication and accentuate the overall superficiality of the systems tasked to emulate human exchange.

Ross Goodwin and Oscar Sharp used this type of slippage to disrupt the cinematic stereotypes in their short film *Sunspring* (2016) [48]. Trained with the 1980s and 1990s sci-fi movie screenplays found on the Internet, Goodwin's ML software generated the screenplay and the directions for Sharp to produce *Sunspring*. It brims with awkward lines and plot inconsistencies but qualified with the top ten entries in the Sci-Fi London film festival's 48-Hour Film Challenge. *Sunspring* reverses the corporate movie search algorithms and playfully mimics contemporary Hollywood's screenwriting strategies largely based on regurgitating successful themes and narratives from earlier films [2] (pp. 390–392). By regurgitating *Sunspring*'s concept and methodology two years later, Alexander Reben produced "the world's first TED talk written by an A.I. and presented by a cyborg" *Five Dollars Can Save the Planet* (2018) [49]. A YouTube comment by MTiffany fairly deems it "Just as coherent, relevant, and informative as any other TED talk." [50].

Another approach to implicit uncanniness is by alluding the intimate familiarity of the human body, for example in Scott Eaton's *Entangled II* (2019) [51] which is comparable to earlier, structurally more sophisticated video works such as Gina Czarnecki's *Nascent* and *Spine* (both 2006) [52], or Kurt Hentschläger's *CLUSTER* (2009–2010) and *HIVE* (2011) [53]. Ironically, projects that combine uncanniness with our apophenic perception in order to "humanize" AI often contribute to diverting attention from pertinent socio-political issues. For example, with *JFK Unsilenced: The Greatest Speech Never Made* (2018, commissioned by the Times) [54], the Rothco agency aimed at contemplative uncanniness by exploiting the emotional impact of sound to reference the romanticized image of John F. Kennedy. Based upon the analysis of 831 recorded speeches and interviews, Kennedy's voice was deepfaked in a delivery of his address planned for the Dallas Trade Mart on 22 November 1963. The voice sounds familiar at the level of individual words and short phrases, but its overall tone is uneven, so the uncanniness relies mainly on the context of the speech that the young president never had a chance to give. However, even with perfect emulation of accent and vocalization, this exercise could never come close to matching the eeriness and deeply disturbing political context of Kennedy's televised speech on 22 October 1962 about the Cuban missile crisis in which sheer good luck prevented the multilateral confusion, incompetence, ignorance, and insanity of principal human actors from pushing the world into nuclear disaster [55].

Visual deepfakes, such as Mario Klingemann's *Alternative Face* (2017) [56] or Libby Heaney's *Resurrection (TOTB)* (2019, discussed in Section 2.4.1), approach the psycho-perspective mechanism of uncanniness explicitly, by simultaneously emphasizing and betraying the visual persuasiveness of statistically rendered human-like forms. This strategy was prefigured conceptually and procedurally by Sven König's *sCrAmBlEd?HaCkZ!* (2006) [57] which facilitated continuous audiovisual synthesis from an arbitrary sample pool. It used psychoacoustic techniques to calculate the spectral signatures of the audio subsamples from stored video material, and saved them in a multidimensional database; the database was searchable in real-time to mimic any sound input by playing the matching audio subsamples synchronized with their corresponding video snippets. Perhaps this innovative project has been largely forgotten because König pitched it to the VJ scene rather than using it to develop artworks that establish meaningful relations between their stored videos and input audio (both selectable by the artist). Along with the sophistication of his technique, König's expressive mismatch may have anticipated the analogous issues in contemporary AI art.

2.1.4. The Mechanical Turkness

The socio-political aspects of anthropomorphism can be effectively addressed by addressing the deep social embeddedness of complex technologies such as AI and by exposing human roles and forms of labor behind the "agency" or performative efficacy of corporate AI.

For example, Derek Curry and Jennifer Gradecki's project *Crowd-Sourced Intelligence Agency (CSIA)* (since 2015) [58] offers a vivid educational journey through problems, assumptions, or oversights inherent with ML-powered dataveillance practices. It centers around an online app that partially replicates an Open Source Intelligence (OSINT) system, and allows the visitors to assume the role of data security analysts by monitoring and analyzing their friends' Twitter messages, or by testing the "delicacy" of their own messages before posting them. The app features an automated Bayesian classifier designed by the artists and a crowdsourced classifier trained on a participant-labeled data from over 14,000 tweets, which improves its accuracy by the visitors' feedback on its previous outputs. *CSIA* also includes a library of public resources about the analytic and decision-making processes of intelligence agencies: tech manuals, research reports, academic papers, leaked documents, and Freedom of Information Act files [59]. This multilayered relational architecture offers an active learning experience enhanced by the transgressive affects of playful

"policing" in order to see how the decontextualization of metadata and the inherent ML inaccuracies can distort our judgment.

Similarly, RyBN and Marie Lechner's project *Human Computers* (2016–2019) [60] provides revelatory counter-intuitive insights into the use of human beings as micro-components of large computational architectures. It is based upon a multi-layered media archaeology of human labor in computation since the 18th century. It shows that many AI applications have in fact been simulacra, mostly operated by echelons of underpaid workers, which corporate AI euphemistically calls "artificial Artificial Intelligence" (AAI) or "pseudo-AI". This foundational cynicism of corporate AI indicates that its development imposes an exploitative framework of cybernetic labor management [20,61], which significantly diverges from Norbert Wiener's cybernetic humanism in *The Human Use of Human Beings* (1988) [62].

A sub-project of *Human Computers*, titled *AAI Chess* (2018), was an online chess app with three all-human playing modes: human vs. human, human vs. Amazon MTurker, and MTurker vs. MTurker. In 2020, Jeff Thompson "replayed" *AAI Chess* with his performance *Human Computers* [63], in which the visitors were tasked to manually resolve a digital image file (Google StreetView screenshot of the gallery) from its binary form into a grid of pixels. With 67 calculations per pixel, the complete human-powered image assembly takes approximately eight hours. Here, the visitors' unmediated enactment of automated operations asserts how a combination of complexity and speed in pervasive technologies makes them difficult to understand and manage by an individual.

These projects were conceptually and methodologically anticipated by several earlier works, particularly by Kyle McDonald and Matt Mets' *Blind Self Portrait*, and Matt Richardson's *Descriptive Camera* (both 2012) [64,65]. In *Blind Self Portrait*, a laptop-based face recognition setup draws linear portraits of the visitors, but in order for the setup to work, the sitter has to keep eyes closed while resting their hand on a horizontally moving platform and holding a pen on paper. Unlike Van Arman's, Hastie's, or Patrick Tresset's drawing robots (*Human Studies* series since 2011) [66], which put their sitters in a traditionally passive role, *Blind Self Portrait* makes a reference (intended or not) to William Anastasi's *Subway Drawings* from the 1960s [67] and playfully turns visitors into the "mechanical parts" of a drawing system, self-conscious of their slight unreliability. Richardson's *Descriptive Camera* has a lens but no display; it sends the photographed image directly to an Amazon MTurker tasked to write down and upload its brief description, which the device prints out.

By exploiting human labor in order to emulate the features of AI systems or AI-enabled devices, these projects remind us that the "Turk" in AI is still not mechanical or artificial enough, he resists "emancipation", and it is not easy to make him more "autonomous". Their self-referential critique also points to the ethically questionable use of non-transparent crowdsourcing in art practices exemplified by earlier Aaron Koblin's projects *The Sheep Market* (2006), *10,000 Cents* (2008), and *Bicycle Built for Two Thousand* (2009, with Daniel Massey) [14] (pp. 117–120) [68].

It is noteworthy, however, that artistic attempts to approach computational creativity through active open-sourced participation can be equally undermined by the muddled relationship with anthropomorphic notions. Seeing ML as a tool that "captures our shared cognitive endowments", "collective unconscious", or "collective imagination" [69], Gene Kogan initiated a crowdsourced ML project *Abraham* in 2019 with a goal to redefine agency, autonomy, authenticity, and originality in computational art. The opening two parts of Kogan's introductory essay describe *Abraham* as "an open project to create an autonomous artificial artist, a decentralized AI who generates art", and elaborate on the idea in a semantically correct but conceptually derisive discussion, raising suspicion that the author is unaware of Jaron Lanier's prescient critique of online collective creativity and the subsequent relevant work [70–72]. The missing parts 3 and 4 of the essay were proposed to be published by the end of 2019 [73].

*2.2. Epistemological Space*

Art methodologies that address the epistemological character and boundaries of ML often involve sampling of multi-dimensional datasets in the inner (hidden) layers of network architectures and rendering their representations compressed in two or three dimensions. Artists treat these datasets as a latent space, a realm between "reality" and "imagination", replete with suggestions that emerge from a complex interplay between the various levels of statistical abstraction or determination [21] (p. 9).

2.2.1. Inceptionism

The exploration of latent space started with adaptations of a CV software package DeepDream in order to produce imagery and animations in a quasi-style called Inceptionism, characterized by delirious fractal transformations of pareidolic chimeras [74]. Examples include Mike Tyka's *DeepDream* (2015–2016); Gene Kogan's *DeepDream Prototypes* (2015); James Roberts' *Grocery Trip* (2015); Samim Winiger's *ForrestGumpSlug* (2015); Josh Nimoy's *Fractal Mountains*, *Hills*, and *Wall of Faces* (all 2015); Memo Akten's *All Watched Over* and *Journey Through the Layers of the Mind* (both 2015); Johan Nordberg's *Inside an Artificial Brain* (2015) and *Inside Another Artificial Brain* (2016). Inceptionist works struggled to become more than decorative interventions, and the trend dried out relatively quickly. Besides the inherent structural uniformity and apparent formal similarities between Inceptionist works, the main reason is that arbitrary generation of mimetic imagery or animations tends to become oversaturating and boring if it unfolds unbounded. In order to engage the viewer, it requires prudently defined conceptual, narrative, and formal constraints, which seemed to be difficult to implement with DeepDream.

2.2.2. Sampling the Latent Space

Further experimentation prompted artists to transcend mere representation by exploiting ML with meaningful premises, and by finding more flexible aesthetics to mediate the latent space. To metaphorize the statistically structured epistemological scope of neural network architectures, artists often accentuate the tensions between their processual effectiveness and interpretative limitations.

Timo Arnall's *Robot Readable World* (2012) [75] is an early example of this approach. It comprises found online footage of various CV and video analytics systems (vehicle and crowd tracking, counting and classification, eye-tracking, face detection/tracking, etc.), composited with layers that visualize their data in real-time. However, Arnall's attempt to reveal the "machinic perspectives" uses a human-readable (anthropocentric) approximation of the actual software data processing, tracing back to the funny but technically groundless translation of the terminator android's CV data to English in *Terminator 2* (1991, directed by James Cameron). In *Computers Watching Movies* (2013) [76], Ben Grosser handled this topic more appropriately. It illustrates the CV processing of six popular film sequences in a series of temporal sketches in which the points and vectors of the CV's focal interest are animated as simple dots and lines on a blank background (the processed film footage is not visible), synchronized with the original film sound. This semi-abstraction draws viewers to make sonically-guided comparisons between their culturally developed ways of looking and the "attention" logic of CV software that has no narrative or historical patterns.

In contextually different settings, the semantic power of written text can provide strong generative experiences. For example, Nao Tokui and Shoya Dozono's *The Latent Future* (2017) [77] is an ambient installation based on the interaction between an ML semantic model trained on a collection of past news and the real-time human or machine-generated news. It continuously captures Twitter newsfeeds and uses their discerned meanings to create fictional news. The generated news is presented in a virtual 3D space that maps each sentence's latent feature vectors, while the distances between the sentences correspond to their relative semantic differences. This work is informed in real-time by the largely unpredictable dynamics of the Twitter galaxy, but also by the Twitter's filtering algorithm which represents many important aspects of current socio-political trends.

The approach to interpretation in these and many other AI artworks calls for a comparison with earlier generative works which reveal the tropes of various media in aesthetically elegant and intellectually engaging ways [3]. For example, in Memo Akten's *Learning to See* (since 2017) [78] visitors are invited to arrange various household objects on a table for a camera feed that is processed in real-time by a convolutional conditional generative adversarial network (GAN) autoencoder which mimics the input shapes and surface patterns as compositions of clouds, waves, fire bursts, or flowers, depending on the chosen training model. By revealing narrowness and arbitrariness, the ambiguous interpretative efficacy of this interaction suggests the similarity between GAN's and human vision in their reliance on memory and experience. However, the actual experience of *Learning to See* quickly becomes tedious and erodes into a mildly amusing demo because, regardless of the object arrangements or the selection of interpretative image dataset, the results are always homogenously unsurprising.

The relational flexibility of human visual interpretation that *Learning to See* fails to address was brilliantly utilized by Perry Bard in a conceptually and formally comparable non-AI generative project *Man with a Movie Camera: The Global Remake* (2007–2014) [79]. It is an online platform that allows visitors to select any shot from Dziga Vertov's seminal film *Man with a Movie Camera* (1929) and upload their video interpretations. Bard's server-side software replays a two-channel setup comprising Vertov's original synchronized with a remake which is continuously assembled of the participants' shots (randomly selected when there are multiple uploaded interpretations of the original shot). By leveraging the creative breadth of human perception and cognition, this relatively simple technical setup engrosses both the uploaders and viewers in an intriguing and surprising experience.

2.2.3. GANism

In order to explore and mediate the latent space, artists have been developing various techniques that exploit the increasingly versatile GAN architectures. They reveal the artifactual character of GANs by treating autoencoder networks as compression algorithms, for example in Terence Broad's *Blade Runner—Autoencoded* (2016) [80], or by allowing "unpolished" representations of GAN data, for example in Elle O'Brien's *Generative Adversarial Network Self-Portrait* (2019) [81], or Jukka Hautamäki's *New Parliament* (2019) and *Restituo I and II* (2021) [82,83].

Other examples of manipulating the latent space include Anil Bawa-Cavia's *Long Short Term Memory* (2017); Gene Kogan's *WikiArt GAN* and *BigGAN Imitation* (both 2018); several Mario Klingemann's portrait synthesis works, such as *Face Feedback*, *Freeda Beast* (both 2017), or the *Neural Glitch* series (2018) [56]; AI Told Me's *What I Saw Before the Darkness* (2019); Hector Rodriguez's *Errant: The Kinetic Propensity of Images* (2019); Sukanya Aneja's *The Third AI* (2019); Tasos Asonitis' *Latent Spaces* (2021), and others. Within this range of works, Weidi Zhang's *LAVIN* (2018) [84] is notable for its sampling/rendering methodology and representational strategy. It provides a responsive virtual reality (VR) experience of a GAN which maps the real-world objects from a video camera feed to the semantic interpretations limited to a set of less than a hundred daily objects; the photogrammetric reconstructions of these objects navigate the audience through a virtual world.

Recent GAN techniques allow complex formal remixing by modifying generator or discriminator networks, for example in Golan Levin and Lingdong Huang's *Ambigrammatic Figures* (2020) [85], or by inserting filters and manipulating activations maps in the higher network layers to disrupt the image formation process, for example in Terence Broad's *Teratome* (2020) [86].

Due to their limited autonomy in choosing the training datasets or the statistical models that represent the latent space, GANs prove to be primarily the tools for processual mimicry rather than intelligent creative engines [21] (p. 9). Therefore, GAN manipulation renders ubiquitous visual character in disparate works produced with similar techniques. Although some projects go beyond purely technical/formal exploration or perceptual study, aesthetically (and often conceptually), many do not diverge significantly from earlier glitch art in which the error is an aestheticized frontline layer [87]. This expressive issue reaffirms the importance of the artist's decision-making and overall poetic articulation. In contrast to the tech community's quick approval [88] and self-conceited assertions that "GAN artists have successfully cultivated their moderately abstract, dream-like aesthetic and promoted the process of serendipitous, often random usage of generative processes" [89], the poetic identity of GAN artworks is dominated by Dali-esque or Tanguy-esque formal fusion (morphing), often visually oversaturated but conceptually bland. Superficiality and banal consumerist notions of perception in GAN art extend to the socio-political and ethical dimensions by pointing to the artists' technocratic strategies which Żylińska critically labels as "platform art" [14] (pp. 77–85).

The popularity of GANs has also escalated the misuse of the expression "generative art" to describe only the computational practices that involve randomness, complexity, or ML architectures. A disregard for the methodological diversity and long history of generative art [90,91] impoverishes the broader contextual milieu of experimental art and facilitates the uncritical appreciation of AI art practices.

*2.3. Spectacularization*

AI art with a highest public visibility profile comprises derivative projects by mainstream artists, and big-budget AI art spectacles.

2.3.1. Derivative

The AI's growing ideological authority, socio-economic power, and the practical accessibility of ML software had induced the MCA's involvement with AI art in the mid-2010s. The recent adoption of blockchain crypto products such as non-fungible tokens (NFTs) for securing the marketability of digital entities further increased the gallery/museum and auction house interest [92], prompting the mainstream artists to assimilate ML into their repertoire and to update their poetic rhetoric accordingly. Similar to the post-digital artists a decade earlier [93], they approach digital technologies as affective markers of contemporary culture and act chiefly in collaboration with hands-on personnel to produce AI-derived works in conventional media (installation, sculpture, video, and photography) with a lower degree of technological entanglement than most experimental AI artworks.

This strategy affords them cultural recognition, institutional support, and commercial success, but sacrifices the intricate tension between the artworks' conceptual, expressive, or narrative layers and the contextual logic of the technologies in which they appear. Examples include Gillian Wearing's *Wearing Gillian* (2018); Lucy McRae's *Biometric Mirror* (2018); Hito Steyerl's *Power Plants* and *This is the Future* (both 2019); Pierre Huyghe's *Of Ideal* (since 2019) and *UUmwelt* (2019); Kate Crawford and Trevor Paglen's *Excavating AI* (2018) and *Training Humans* (2019–2020), and others. The presentational authority and decorative appeal of this production tend to seduce the general audience into superficial aesthetic consumption or complacency, even when projects are created with critical intentions.

For example, Trevor Paglen's AI-related production has been praised as a critique of biases, flaws, and misconceptions of AI technologies, along with his established line of interest in visualizing the covert systems of power and control in the military, intelligence, state, or corporate institutions. However, it is also criticizable as an exploitation of activist perspective toward opto-centric epistemology, which mystifies high-end visual technologies and abuses the affective perception of institutional power through stylized gallery setups accompanied with highfalutin explanatory statements. Paglen's collaborative project with Kate Crawford *ImageNet Roulette* (2019) [94] is represented as being critical of classification biases in CV, but it is hard to see in it anything more than an overwhelming illustration of the issue. As an analytical, research-based revelatory critique of classification biases and AI technologies in general, it is neither new nor original, for example when contrasted with Curry and Gradecki's *CSIA* (since 2015, discussed in Section 2.1.4) or with RyBN's systematic critical analysis in a number of projects, such as *Antidatamining* (since 2007) [95] and *Human Computers* (2016–2019, discussed in Section 2.1.4) [60]. Similarly, its socio-cultural commentary fades in comparison with Taryn Simon and Aaron Schwartz's project *Image Atlas* (since 2012) [96] which addresses the same issues, but discards loftiness for a more meaningful impact by obtaining simple imagery in complex ways and by coupling it with concise, unpretentious narratives.

2.3.2. Large Scale

The substantially market-driven operational criteria and depoliticized discourses of MCA [97,98] have been epitomized by spectacular, large-scale AI art installations in various forms: static, generative, reactive, interactive, or self-modifying "intelligent environments" [17]. This high-profile/high-visibility approach had been ushered with corporate enterprises such as *The Next Rembrandt* (2016) [99], collaboratively produced by ING bank, Microsoft, Technical University in Delft, and Mauritshuis art collection. They used DL for a complex multi-feature analysis of Rembrandt's paintings in order to generate and 3D print a "most representative" painting of his style. The project's promo language is typical of the corporate AI's patronizing anthropomorphism, claiming that it "brought the great master back to life".

Examples of large-scale AI art installations include Sosolimited, Plebian Design, and Hypersonic's *Diffusion Choir* (2016); Marco Brambilla's *Nude Descending a Staircase No. 3* (2019) [100]; Refik Anadol studio's projects such as *Melting Memories* (2017), *Machine Hallucination* (2019 and 2020), and *Quantum Memories* (2020) [101]; CDV Lab's *Portraits of No One* (2020) [102]; projects by Ouchhh studio; projects by Metacreation Lab, and others.

Along with many GAN works discussed in Section 2.2, these practices willingly or unwillingly contribute to platform aesthetics—a mildly-amusing algorithmic generation of sonic, visual, spatial, or kinetic variations, which teases the visitors with the promise of novelty and insight, but effectively entrances them into cultural conformity and political deference [14] (pp. 72–83, 132–133). Dependent on the latest AI research and elaborately team-produced with significant budgets or commissions, the hyper-aestheticized AI art installations also warn how effectively the manipulative intents, unimpressive concepts, or trivial topics can be concealed behind skillful rendering, aggrandized by high production values, and popularized through flamboyant exhibition.

The issues of platform aesthetics are exemplified by the AI installations produced in Refik Anadol's studio [101], which flirt with sophisticated production techniques, formal oversaturation, and inflated presentation. Their dubious motivations are clumsily veiled by inane flowery premises and by infantile anthropomorphic metaphors such as "transcoding the processes of how buildings think", or "how AI systems dream" or "hallucinate". Anadol has frequently claimed his childhood fascination with the spectacular advertising in *Blade Runner* (1982, directed by Ridley Scott) as one of the uplifting inspirations for his art career, without any self-critical reevaluation of the political background of visuals and architecture in that film. Only in 2021, he was induced to acknowledge his misreading of the dystopian essence of *Blade Runner*'s aesthetics [103]. Consequently, despite the formal abundance and copious explanatory data (which usually do the opposite of demystifying the production), Anadol's spectacles have been virtually devoid of critical views on mass surveillance, immaterial labor, environmental damage, and other problematic aspects of the big data capture and processing they rely upon. For a comparison, we can take some of the monumental art practices throughout the 1980s that roughly coincided with the release of *Blade Runner*, such as Krzysztof Wodiczko's projections [104], Barbara Kruger's immersive setups [105], or Anselm Kiefer's heavy confrontational installations [106]. They employed grand scale, formal saturation, and overidentification to critically appropriate and reflect the inherent use of overwhelming presentational strategies in gender-biased advertising, power-structures, and totalitarian regimes. While the tactical values of these practices had been thereafter attenuated or recuperated in an inevitable process of cultural assimilation, they redefined the landscape of critical art with lasting historical impact and relevance.

Another telling parallel can be drawn between Marco Brambilla's *Nude Descending a Staircase No. 3* (2019) [100] and Vladimir Todorović's *The Running Nude* (2018) [107], which both relate to Marcel Duchamp's painting *Nude Descending a Staircase No. 2* (1912). Brambilla relies on compositional abundance and installation size to sustain the GAN animation that refers to the influence of early cinema on cubism and futurism. Todorović's generative VR work unobtrusively leverages the problem of data interpretation in AI to reference the polyvalent interpretation in western fine arts tradition. It provides a formally subdued but experientially intensive interactive experience in which the ASMR-whispered descriptions of select classical nude paintings are generated by an ML program trained on pulp love stories.

To extend the exploration of this sweeping comparative range, the reader is invited to relate the spectacular generative portrait synthesis in CDV Lab's installation *Portraits of No One* (2020) [102] with formally compact and technologically discreet works such as Jason Salavon's *The Class of 1967 and 1988* (1998) [108]; Golan Levin and Zachary Lieberman's *Reface (Portrait Sequencer)* (2007–2010) [109]; or Shinseungback Kimyonghun's *Portrait* (2013) [110].

The crass complacency exerted by spectacular AI art suggests that its creators have skipped some of the required reading assignments of Modern Art History 101 courses, most notably Guy Debord's *The Society of the Spectacle* (1994) [111]. It discredits the self-serving claims of some cultural agents that spectacular AI art opens up opaque ML technologies, makes them more accessible to the public and thus more exposed to critical assessment [112]. As evident in performative AI art practices discussed in Section 2.1.2, and from the long history of religious art, totalitarian art, or advertising, the aesthetic and presentational exuberance undermine the exploratory and epistemological impact, or conceal the lack thereof. The cultural momentum of uncritical or manipulative AI art spectacles and AI-derived mainstream art is particularly detrimental to the field because it obscures experimental and avant-garde practices, and tempts the emerging AI artists to soften their critical edge in favor of career-friendly strategies [6] (pp. 252–254) [23].

*2.4. Tactical Exploration*

The recurrence of tactical AI art exemplars throughout this study indicates their potential to direct the field toward a socially responsible and epistemologically relevant expressive stratum. Tactical AI art extends the heterogeneous line of critical practices in new media art, which have energized art and culture in the 20th and the 21st century by subverting and exposing the exploitative corporate strategies based on quantization, statistical reductionism, data-mining, behavioral tracking, prediction, and manipulation of decision-making [3] (pp. 71–73). Artists uncover the undesirable aspects and consequences of corporate AI and denounce biases, prejudices, economic inequalities, and political agendas encoded in the mainstream ML architectures. In some works, they also engage in an exploratory critique of the nature of ML as an artistic medium; the value of this critique is proportional to the artists' understanding of the political subtleties and ethical facets which are often dispersed across the conceptually abstract, technically convoluted, and functionally opaque ML systems.

To incite active critical scrutiny, artists sometimes combine humor and provocation by intentionally taking seemingly ambivalent positions toward the issues they address; they emulate the corporate AI's operative models but recontextualize them or repurpose their objectives for ironic revelatory effects. One of the common repurposing methodologies involves taking an existent ML pipeline, training it with a nonstandard dataset, and employing it for novel tasks. Many successful tactical works refrain from dramatic interventions and didactic explanations in order to let the audience actively identify the interests, animosities, struggles, inequalities, and injustices of corporate AI.

2.4.1. Socio-Cultural

Artists often work with NLP to critique various cultural manifestations of applied AI. Examples include Matt Richardson's *Descriptive Camera* (2012, discussed in Section 2.1.4); Ross Goodwin's *Text Clock* (2014) and *word.camera* (2015); Michel Erler's *Deep Learning*

*Kubrick* (2016); Ross Goodwin and Oscar Sharp's *Sunspring* (2016, discussed in Section 2.1.3); Jonas Eltes' *Lost in Computation* (2017, discussed in Section 2.1.3); Jonas Lund's *Talk to Me* (2017–2019); Joel Swanson's *Codependent Algorithms* (2018), and others.

A number of related projects use NLP and language hacking to probe the intersection of AI technologies and MCA. For example, Disnovation.org's *Predictive Art Bot* (since 2017) [113] questions the discursive authorities and aesthetic paradigms of AI art; Sofian Audry and Monty Cantsin's *The Sense of Neoism!* (2018) critiques the cogency of artists' manifestos and proclamations; Philipp Schmitt's *Computed Curation Generator* (2017) and Alexander Reben's *AI Am I (The New Aesthetic)* (2020) problematize art-historical models and narratives; Nirav Beni's *AI Spy* (2020) and Egor Kraft's *Museum of Synthetic History* (2021) address culturally entrenched aesthetic paradigms.

For interventions that relate to the socio-cultural issues of AI, artists use and modify GAN architectures to make deepfakes. For example, Libby Heaney's *Resurrection (TOTB)* (2019) [114] thematizes both the star-power in music and the memetic power of deepfakes. Visitors of this installation are invited to perform karaoke in which the original musician of the chosen song is video-deepfaked to mimic the visitor's singing and gesturing/dancing. Additionally, in between songs the host Sammy James Britten involves the audience in the discussion of power, desire, and control—an extension that seems to be as imposing and redundant as the artist's explanatory section for this work. Heaney's *Euro(re)vision* (2019) [115] addresses the transmission of power and politics in popular media more effectively. In this video deepfake, Angela Merkel and Theresa May sing absurd songs in the style of Dadaist Cabaret Voltaire performances within a setting of the Eurovision song contest. Their stuttering algorithmic poetry eerily resembles the nonsensicality of actual Brexit discourse and implies the broader semantic reality of political life.

With two iterations of *Big Dada: Public Faces* (2019–2021) [116], Bill Posters and Daniel Howe confused the visitors of Instagram by inserting deepfaked fictional video statements by Marcel Duchamp (about the ashes of Dada), Marina Abramović (about mimetic evolution), Mark Zuckerberg (about the second Enlightenment), Kim Kardashian (about psycho-politics), Morgan Freeman (about smart power), and Donald Trump (about truth).

In several works, Jake Elwes critically engages the cultural implications of training dataset annotation and algorithm design in mainstream AI. His ongoing multipart *Zizi Project* (since 2019) [117] interfaces deepfake with the world of LGBTQ+. *Zizi-Queering the Dataset* (2019) is a video installation continuously morphing through gender-fluid (androgynous) portraits and abstract forms. The online work *Zizi Show* (2020) critiques both anthropomorphism and error-prone gender inclusiveness of AI. This virtual drag cabaret features deepfakes generated from the training datasets with original films of London drag artists' performances. The *Zizi Project* clearly indicates that the training model datasets and statistical nature of data processing in GANs inevitably impose formal constraints to the possible outputs (such as realistic human-like images) regardless of the common rhetoric about the "unpredictability" or "originality" of such systems; however, this is an already known and well-documented issue [10] (pp. 9–10). The project fails to show how exactly the race, gender, and class inequalities and stereotypes transfer into ML to harm the underrepresented social, ethnic, or gender identity groups. The *Zizi Project*'s playful, technically sophisticated remediation within AI-influenced cultural context may be beneficial for the celebration, affirmation, and inclusion of LGBTQ+, but its publicity narratives, its high production values, and its focus on glamour and spectacle in lieu of less picturesque but perhaps more important existential aspects of LGBTQ+ can easily be perceived as artistic exploitation by means of ML. Moreover, if taken seriously by corporate AI, this critique can backfire by contributing to the refined normalization, instead of correction, of socio-political biases toward the LGBTQ+ community because these biases have a broader, deeper, and darker evolutionary background.

In contrast, Derek Curry and Jennifer Gradecki's *Infodemic* (2020) [118] and *Going Viral* (2020–2021) [119] exemplify a consistently more effective critique, recontextualization, and transformation of ML as a socio-technical realm [59]. Both projects target celebrities, influ-encers, politicians, and tech moguls who have "contributed" to the COVID-19 pandemic by sharing misinformation and conspiracy theories about the coronavirus, which themselves went "viral", often spreading faster than real news. *Infodemic* features a cGAN-deepfaked talking head video in which some of these high profile mis-informers deliver public service announcements which correct false narratives about the pandemic; their statements are taken from and voiced by academics, medical experts, and journalists. In *Going Viral*, visitors of the project website are invited to help intervene in the infodemic by sharing on social media the "corrective" videos delivered by the deepfaked speakers in the *Infodemic*. By playing with deepfakes within their native context of fake news, these projects also probe the broader phenomenology of mediated narratives. Together with *CSIA* (discussed in Section 2.1.4), they testify to the effectiveness of Curry and Gradecki's tactics based on thorough research and self-referential methodology with computational media affordances. A specific quality of their poetics is that playful participation is simultaneously a gateway to transgressive affects, an interface to learning resources, and a friendly implication of our complicity to the politically problematic aspects of the applied AI through conformity, lack of involvement, or non-action.

2.4.2. Physical and Existential

AI technologies affect socio-cultural life and politics both directly and indirectly, through the material/physical, ecological, and existential changes. Artists sometimes metaphorize this influence by using geospatial contents (landscapes, terrains, maps) for training datasets and by positioning the machine-learned output in contexts with various political connotations.

For example, Ryo Ikeshiro's *bug* (2021) [120] is a sophisticated geospatial ambient work that addresses the uses of ML-powered sound event recognition and spatial/directional audio technologies in entertainment, advertising, surveillance, law enforcement, and the military. Similarly, Nao Tokui's *Imaginary Landscape* and *Imaginary Soundwalk* (both 2018) [77] are formally economical interactive installations. In *Imaginary Landscape*, the ML software continuously analyzes Google StreetView photographs, selects three that look similar, and joins them together horizontally in a three-wall projection. Another ML program, trained on landscape videos, generates soundscapes that correspond with stitched triptych landscapes. In *Imaginary Soundwalk*, viewers freely navigate Google StreetView for which the ML system, using the cross-modal technique for image-to-audio information retrieval, generates the "appropriate" soundscape. It is instructive to compare the meditative effectiveness of these projects with Anna Ridler and Caroline Sinders' interactive online work *Mechanized Cacophonies* (2021) [121].

Other examples include Mike Tyka's *EONS* (2019); Liliana Farber's *Terram in Aspectu* (2019); Weili Shi's *Martian Earth* and *Terra Mars* (both 2019); Martin Disley's *the dataset is not the map is not the territory* (2020), and Daniel Shanken's *Machine Visions* (2022).

Some works explore the physicality of AI through haptics (touch), for example Jeff Thompson's *I Touch You and You Touch Me* (2016–2017) [122], or through kinetics, for example Stephen Kelly's lumino-sonic installation *Open Ended Ensemble (Competitive Coevolution)* (2016) [123]. François Quévillon's *Algorithmic Drive* (2018–2019) [124] also uses kinetics to play out the tension between robotics and the unpredictable nature of the world. For this work, several months-worth of front facing video capture was synchronized with information from the car's onboard computer, such as geolocation, orientation, speed, engine RPM, stability, and temperatures at various sensors. The captured videos and data feed a sampling system that sorts the content statistically and assembles a video that alternates between calm and agitated states by modifying parameters of sound, image, car's activity, and environment. An interactive controller displays data for each scene and allows visitor intervention.

Continuing the line of earlier statistically founded eco-conscious tactical media art, such as Chris Jordan's *Running the Numbers* (since 2006) [125], artists combine speculative approach with ML to generate visuals and narratives that address the environmental challenges and ecological aspects of large-scale computation-intense research, technologies, and industries such as AI. Examples include Tivon Rice's *Models for Environmental Literacy* (2020) [126], and Tega Brain, Julian Oliver, and Bengt Sjölén's *Asunder* (2021) [127]. Maja Petrić's *Lost Skies* (2017) [128] illustrates how much easier it is for the projects in this range to aestheticize the ecological data than to articulate it into meaningful and perhaps actionable narratives. Ben Snell's *Inheritance* (2020) [129] elegantly and somewhat provocatively compresses the material and ecological aspects of AI. It is a series of AI-generated sculptures cast in the composite medium which was produced by pulverizing the computers used to generate the sculptures' 3D models. This project also addresses the issues of agency and creative expression by referencing radical auto-recursive art experiments such as Jean Tinguely's self-destructive machines. Expectedly, regardless of their poetic values, it is not easy to calculate how much the systemic technological entanglements of such projects (and AI art in general) participate in the overall environmental damage and contribute to the legacy of the Anthropocene.

A full spectrum of the applied AI's existential consequences is boldly integrated in Max Hawkins' *Randomized Living* (2015–2017) [130]. In this two-year experiment, Hawkins organized his life according to the dictate of recommendation algorithms. He designed a series of apps that shaped his life by randomized suggestions based on the online data: a city where he would live for about a month and, once there, the places to go, people to meet, and things to do. *Randomized Living* is a strong exemplar of cybernetic-existentialism—the art of conceiving a responsive and evolving cybernetic system in order to express deep existential concerns [42].

2.4.3. Political

The uneasy positioning of the individual toward or within computational systems of control has been reverse-engineered in a number of works by new media artists and activists such as Bureau d'Etudes, Joana Moll, Adam Harvey, and Vladan Joler. In several collaborative projects, Joler has been effectively applying analytical tools and mapmaking to render diagrams of AI power within various perspectives. With SHARE Lab and Kate Crawford, he released *Exploitation Forensics* (2017) [131] which snapshots in a series of intricate diagrams the functional logic of Internet infrastructure: from network topologies and the architecture of social media (Facebook) to the production, consumption, and revenue generation complex on Amazon.com. Similarly, Crawford and Joler's project *Anatomy of an AI System* (2018) [132] deconstructs the Amazon Echo device's black box by mapping its components onto the frameworks of global ecology and economy. With Matteo Pasquinelli, Joler issued *The Nooscope Manifested* (2020) [133], a visual essay about the conceptual, structural, and functional logic of sub-symbolic ML, and its broader epistemological and political implications. It leverages the notions of gaze and vision-enhancing instruments as metaphorical and comparative devices, although their conceptual suitability within the context of ML is somewhat unclear.

Since the introduction of the OpenCV library in 2000, artists have been using CV for various purposes in a large corpus of works. With advances in ML, this exploration has intensified and increasingly involved the critique of the (ab)use of CV for taxonomic imaging, object detection, face recognition, and emotion classification in info-capitalism. For example, Jake Elwes' video *Machine Learning Porn* (2016) [134] indicates human (perceptive) prejudices that influence the design of ML filters for "inappropriate" content. Elwes took the open_nsfw CNN which was originally trained with Yahoo's model for detecting "sexually explicit" or "offensive" visuals and repurposed its recognition classifiers as parameters for generating new images. This modification outputs visually abstract video frames with a "porny" allusiveness. However, the cogency of this project depends on leaving out that *all* formal image elements are abstract by default and that in humans, the pathways of complex scene recognition and related decision-making are not precisely known [135,136], so the ground for critiquing biases in these pathways is also uncertain.

The issues of ML-powered biometry are particularly sensitive and pertinent in facial recognition and classification due to the convergence of evolutionarily important information in the face and its psycho-social role as the main representation of the self and identity. Various deficiencies frame the machine training/learning and "recognition" process in which the classification models ultimately always make implicit (but unobjective) claims to represent their subjects.

Some critical works in this domain function as markers of the technical improvements in face recognition, for example Zach Blas' *Facial Weaponization Suite* (2011–2014) [137] and *Face Cages* (2015–2016) [138]; Heather Dewey-Hagborg's *How do You See Me* (2019) [139]; and Avital Meshi's *Classification Cube* (2019) [140], or provide demonstrations of expression analysis, for example Coralie Vogelaar's two works in print *Happy*, and *Facial Action Coding System* (both 2018); Lucy McRae's interactive data visualization setup *Biometric Mirror* (2018, mentioned in Section 2.3.1); and Lauren Lee McCarthy and Kyle McDonald's *Vibe Check* (2020). By revealing the human perceptive flaws (such as pareidolia) reflected in CV design, Driessens and Verstappen's *Pareidolia* (2019) [141] reiterates a number of preceding works such as Shinseungback Kimyonghun's *Cloud Face* (2012) and *Portrait* (2013) [110]; Onformative's *Google Faces* (2013) [142]; and Benedikt Groß and Joey Lee's *Aerial Bold* (since 2016) [143].

Biases in ML design have been continuously identified by both scientists and artists. For example, a research project with artistic overtones titled *Gender Shades* (2018) [144] by Joy Buolamwini and Timnit Gebru assessed the accuracy of several corporate facial classifiers (Adience, IBM, Microsoft, and Face++) with respect to gender, skin type, and skin type/gender intersection. Using a custom benchmark dataset with diverse skin types based on 1270 images of parliamentarians from three African and three European countries, Buolamwini and Gebru showed that the error rate of the tested corporate classifiers was significantly higher for women with darker skin color. Their findings affected not only the public but also the US policymakers and the corporate AI sector [145]. Similarly, Kate Crawford and Trevor Paglen's exhibition project *Training Humans* (2019–2020) [146] exposed racial bias in the online image database ImageNet that has been widely used in ML since 2009. Consequently, ImageNet removed 600,000 images of people from its collection of more than 14 million images downloaded from the Internet and annotated by MTurk workers. It is instructive to compare these ID-related AI artworks with Heather Dewey-Hagborg's *Stranger Visions* (2012–2013) [147] which questions the arbitrariness and power games behind the ethics and politics of biometric profiling based on DNA data analysis.

Less didactically structured approaches in this domain allow a wider space for visitors' critical interpretation. For example, Jake Elwes' video installation *Closed Loop* (2017) [148] establishes a mutually generative relational loop between a text-to-image and image-to-text models, which comprises multiple aspects of the CV inaccuracies, biases, and, implicitly, the ethical issues of AI in an unpredictable and witty continuum. Shinseungback Kimyonghun's *Mind* (2019) [110] uses emotion analysis of the last 100 visitors' facial expressions to drive the ocean drums and generate a powerful minimalist sound ambient, with an overhead camera as a single indicator of the machinic gaze. Martin Disley's *How They Met Themselves* (2021) [149] is an open-source project which exploits the recognition borders of face generation/recognition GANs. In a series of steps, it allows visitors to create photorealistic avatars for live webcam deepfaking. Based on the visitor's uploaded portrait, the avatar is created by a generation/discrimination process that yields two visually indistinguishable (virtually identical) images: one is positively identified as the person in the uploaded photo, and the other one is identified negatively (not the person in the photo). The user can upload the generated ambivalent image to train the Avatarify software for a real-time animation of the superimposed avatar in online interactions.

Ironically, unlike the biases in ML, the individual "biases" and creative idiosyncrasies in AI art are desirable but relatively rare. Sebastian Schmieg tackles this deficiency with conceptual relevance, expressive economy, formal clarity, and effectiveness in projects such as *Decision Space* (2016); *This is the Problem, the Solution, the Past and the Future* (2017); *Decisive*

*Camera* (2017–2018); and *Decisive Mirror* (2019) [150]. In different ways, these works feature the unconventional, seemingly absurd, or counter-intuitive taxonomies injected in image classification setups. For example, the visitors of the *Decisive Camera* project website can upload an image which will then be classified within a taxonomic space of four categories: Problem, Solution, Past, and Future, and assigned with a probability percentage for each category. The classification dataset was created in the project's initial phase which invited visitors to select images from the Photographers Gallery's image archive and to assign each image to one of these four categories. This playful subversion places the technical, methodological, and broader socio-political problems of ML design conventions firmly within the human context. It provides the reflections of human nature in the arbitrary authoritarianism of ML classification systems based on the exploitation of human labor for annotating the training datasets.

In projects such as *Myriad (Tulips)* (2018) and *Mosaic Virus: Bitcoin Per Hour* (2018) [151], Anna Ridler critiques the appetite for exploiting the speculative investment strategies wetted by corporate AI and the related crypto technologies. For example, *Mosaic Virus* questions the concepts of ownership, obsessions with wealth, and financial speculation by referring to the historical "tulip mania" phenomenon. Trained on Ridler's custom dataset of roughly 10,000 hand-labelled photographs of tulips, a GAN generates images of tulips inflected by the current Bitcoin values. It links the instability of values projected onto commodified artefacts with the opacity of computational technologies used in creating the work [152]. Benjamin Grosser's online service *Tokenize This* (2021) [76] directs a critical focus onto the commodification of AI art, the hyperproduction, and the rush of speculative transactions on crypto art marketplaces [153]. It humorously subverts the artists' hasty and often uncritical adoption of NFTs, which largely proves to be exploitative and ecologically taxing [92,154].

Since the socio-technical unpredictability is closely related to financial instability, it is worth knowing that AI research, which has been going through successive "springs" and "winters" [7] (pp. 31–32), may end up in Disnovation.org's project *The Museum of Failures* (since 2015) [155]. It is a collection of aborted projects, flops, errors, malfunctions, business failures, ethical rejections, or disasters presented in various formats from historical, symbolic, poetic, and cultural points of view.

**3. Issues**

These examples show that, through success or failure, AI art expands the idea of technologically entangled creativity, and that a conscious consideration of the notion of creativity is a prerequisite for creative endeavors in general. They also point to the human fallacies, cultural constraints, and socio-political ambiguities, which manifest in the conceptual, methodological, ethical, and educational domains of AI art. By identifying, acknowledging, and understanding these issues, artists can refine their creative approaches and find new ways to intervene critically and productively in the AI-influenced social reality.

*3.1. Cogency*

AI research struggles with encoding crucial aspects of human cognition—such as intuition, abstraction, analogy-making, common sense, and inventiveness—into machine intelligence [7] (pp. 200–214) [8] (pp. 160–191). Similarly, the poetic realm of contemporary AI art is most deficient in interesting intuitions, meaningful abstractions, and imaginative analogies. The field particularly lacks projects that use AI systems as means to actualize strong concepts which effectively address the wider perspectives or deeper issues of human existence. Digital technologies offer a generous space for conceptual, as well as formal, methodological, and aesthetic experimentation that can transcend the technologically imposed limits of expression. However, the uneven intellectual breadth and depth, biased or constrained contextual awareness, and sketchy art-historical knowledge affect many AI artists' conceptual thinking, methodologies, and the cogency of their outcomes.

The lack of conceptual sophistication manifests as a disproportion between the artists' computational dexterity, their eloquence in articulating relevant ideas, and their competence with wider artistic, cultural, or historical contexts. Broadly speaking, AI artists and art-related AI researchers tend to ignore a century's worth of artmaking which has moved on from what Marcel Duchamp called the "retinal" paradigm. Meredith Tromble notes that AI art methodologies which use training data comprised of canonical visual imagery resemble western fine arts practices of the 19th century, when painters relied on synthesizing the stylistic elements and aesthetic values of Neoclassicism and Romanticism [19] (p. 4). Artists who make generative AI works sometimes disregard that, in principle, the expressive cogency of generative artworks does not crucially rely on the generative system itself but on the conceptual meaningfulness and economy of the relationship between that system and the artwork's broader context. That reflects a (naïve) lack of understanding that the poetic role of production techniques in the arts fundamentally unfolds and gets emancipated by its coupling with conceptual thinking and contextual awareness. Conversely, artists who exaggerate or fake technical competencies are equally problematic because their works usually miss some interesting technological aspects.

AI art tends to be technologically self-referential as many works rely on tautological or circular concepts based on the artists' ideas about ML technologies. Various notions of bio-detached and socially unembedded creative agency permeate both AI art production and its popular representation through confused, ambiguous, or openly mystifying rhetoric about "machinic artistry". They promote a pseudo-romantic quest for human-flavored creative "essence" within ML systems (and AI in general) instead of demystifying them as socio-political apparatuses which have little to do with creativity per se, and are better understood as sophisticated tools for statistical analysis and measurement [10,16] (p. 6). Complex devices such as computers and software only represent the cumulative human creativity invested in their design, but the artists' self-awareness, reasoning, abstraction, conceptualization, generalization, and analogy-making in dealing with these tools inform the cogency of their works. Their mental abilities, senses, emotions, passions, obsessions, and incentives determine how they interact with the world and make their art. These qualities and aspects should be in the forefront of AI artmaking. Conversely, a responsible approach to AI art requires a clear understanding that—while different forms of creative intelligence are possible and explorable—computers, robots, or algorithms are not artists because they do not embody human social embeddedness, cognitive capabilities, skills, quirks, and, most importantly, human motivations for making art [156–158]. Art is a human dispositive within anthropological and socio-cultural perspectives, so the motivations for expressing creativity through artmaking are partially driven by the evolutionary competitive ambition; among its many functions, art is a socially-constructed system for displaying mating fitness (intelligence, proteanism, wit) and for exhibiting or gaining social status [159,160]. Therefore, it is crucial to acknowledge that the poetic qualities of our artefacts are inherently instrumentalizable as virtue signaling means.

Our unfolding experience with AI confirms that we can learn about technologies (and about the world) equally from their failures and from their successes. ML is based on the assumption that all relevant features can be found within the training data; this is problematic in real life scenarios, especially in experiencing art where the overall context, audience's knowledge, and expectations are essential. ML algorithms can effectively identify frequency-related information about the formal features in artworks and utilize them in increasingly sophisticated ways, but they do not model how an artwork is perceived or interpreted [16] (p. 6). Following that logic, artists use ML to process material from external fields such as visual arts, music, or cinema, aiming to better understand or demystify ML or their processed cultural sources. But designing such ML systems, analyzing gathered information, and deriving meaning from it requires expertise in the logic of extraction, analytical skills, solid knowledge, and deep critical understanding of the processed material. There is no a priori reason to expect that any given AI artist possesses all these attributes because what they ultimately do is artistic experimentation, not the emulation of scientific research. On the other hand, the question is also what special insights they can generate by aestheticizing the artifacts or by spectacularizing the superficial aspects of ML. It is frequently claimed that AI art offers an opportunity for scientists to better understand how machines function [25] (pp. 245, 249). However, a critical view on this claim suggests that, if the scientific community needs art-generated insights to discern the functional logic of its core research subjects (such as deep neural networks), perhaps it is not taking its responsibilities seriously enough. The historical outlook at the disastrous socio-political consequences of applied science/technology [161,162] should remind us that contemporary science, technology, and related businesses need a thorough improvement of epistemological and ethical standards facing the increasing complexity of human existence.

*3.2. Authenticity*

Methodological similarities and aesthetic uniformity of AI art are directly related to the artists' notions of originality. In their production pipelines, artists often use, or sometimes modify, the existing code libraries, and usually train them with commonly available datasets, which leads to homogeneity. Many of them engage in a race to access the emerging code architectures before they become conceptually or aesthetically "exhausted", or to build new training models by curating their own datasets [163]. Some AI artists push the pursuit of technical originality to the brink of obsession [11] (pp. 105, 127) thus revealing an anachronistic fascination with modernist myths of a heroic artist-conqueror. These efforts and priorities also indicate the lack of appreciation that originality is highly contextual, so its overidentification with formal properties is usually misconceived or fetishistic [164].

Production, perception, and reception of the arts have always been evolving in a complex symbiosis with technological and socio-political trends [165], so AI artists—as well as the media and the cultural sector which represent them—should be critically aware of these entanglements. The overall poetics of AI art will remain a facile reflection of its technological reality as long as the majority of artists keep constraining their notions of authenticity and expressive cogency to prima facie relationship with technology. They may benefit from a more general recognition that, in principle, the improved functionality of tech systems such as AI emancipates human intelligence to be expressed in different approaches to artistic creativity. Various forms of apparent but often unacknowledged conceptual or methodological "overlaps", "reflections", and other types of dubious similarities between many artworks demonstrated throughout this paper, indicate the more basic issues of the AI artists' creative literacy and contextual appreciation. These issues are both cognitive and ethical.

*3.3. Technocentrism*

Technocratic or techno-fetishist mentalities have been haunting computational arts since their outset [166,167]. They continue to affect AI art and Żylińska provides a well-grounded critique of the opportunistic and ethically dubious relationship between artists and tech companies such as Google, Amazon, Facebook, Apple, etc. [14] (pp. 75–85).

Successful AI art projects utilize their entanglements self-consciously, as the conceptual, tactical, and existentially inherent features within a broader context of digital culture. However, the complexity, interdependence, and pace of change make digital tools difficult to keep under artistic control. Most notably, the breadth of procedural literacy and coding skills required for elaborate AI art production tend to shape the artists' poetic reasoning, exploration, and learning by directing their creative focus toward mathematics and programming [3] (pp. 75–77). Additionally, the fast competitive pace of producing AI art in current circumstances drains some of the artists' extra energy that comes from idleness and frivolity but often provides an invaluable touch of "dirt" which combines with experimentation, hard work, knowledge, serendipity, luck, and other decisive factors. In general, the engineering approach is usually a welcome enrichment of a "traditional" artistic mindset, but when it takes priority over other poetic factors, it reduces the scope of artists' critical engagement and the impact of their works. When artists are preoccupied with the performative efficiency or with the superficial effects of ML systems, the aestheticized demonstration outweighs discovery in their projects which struggle to engage the audience beyond AI-induced fascination.

Artists are responsible for counteracting these cognitive vectors in their individual working methodologies, but we also need to address them more carefully within academic programs and initiatives which integrate coding skills into a standard learning repertoire of art and design [168]. Although they apparently strive for both versatility and thoroughness [169], most of the creative coding educational environments still tend to be more conducive for nerdy idiosyncrasies than for artistically essential "eccentricities" such as spontaneity, goofiness, wandering, or stubbornness. On the other hand, they sometimes try to promote diversity uncritically or cynically by supporting weak ideas or conceptual naïveté and eventually usher students to adopt subpar professional criteria. Furthermore, these academic frameworks easily migrate from one trendy tech paradigm to another, often leaving the older but still relevant areas underexplored. A recent example is a shift from parametric techniques and symbolic logic to sub-symbolic ML in computational generative art [170].

*3.4. Academism*

Throughout *AI Art* (2020) [14], Żylińska critiques various manifestations of academism in AI art, such as the repetition of topics, similarity of narratives and presentation forms, uniformity of production techniques, and homogeneity of aesthetic models. It is important to expand this critical view by identifying the specter of academism in critically motivated AI art, specifically in the instances when weak concepts, vague reasoning, contextual ignorance, or technophilic self-indulgence are dressed up in a combination of competent execution and emancipatory verbiage. This may be more challenging, but provides valuable insights into the field's undercurrent issues which expose it to recuperation [171] due to the lack of methodological clarity, formal cogency, or experiential impact.

3.4.1. Inflated Speculation

Artists sometimes use speculative forms [172] to stimulate the imagination and explore the "possible worlds" of AI by combining descriptive narratives with illustrative models, props, prototypes, imagery, etc. Speculative artworks leverage both the freedom of thought experiments and the "right to fail" long established by performance art, which claims that—even if it fails to achieve its nominal goal—a performative act may provide a successful experience as long as its context, premise/intent, and process are meaningful. By an unspoken analogy, a speculative project can be legitimate as long as its context, premise/intent, and artefacts are not meaningless, which, of course, does not automatically make the project relevant. Although thinking and language afford incomparably more configurations than the material world, speculative projects are usually less convincing or impactful than artworks that directly actualize the artists' intents or ideas. That is because the speculative freedom and linguistic tolerance allow creators to hedge their expressive stakes; to be venturous but also to conceal the deficiencies of their projects.

In some projects, for example in Tega Brain, Julian Oliver, and Bengt Sjölén's AI-powered environmental speculation *Asunder* (2019, mentioned in Section 2.4.2), the critical logic is so counter-intuitive, its points so sophisticated, and outcomes so delicate that its political vector is sidelined by the discussion about the work [173]. This virtualization of critical focus (or purpose) may be fruitful within the academic milieu, but with a general audience, which is supposedly central to tactical art, it can easily be recognized as aloofness or cynicism and lead to indifference, distrust, or resentment.

3.4.2. Formal Dryness and Compromised Impact

Solidly conceived and well-motivated tactical concepts are sometimes rendered as dry, unengaging, critically ineffective, or counter-effective works. Together with several projects discussed in Section 2, Tom White's *Perception Engines* (2018 and 2021) [174] and Ben Bogard's *Zombie Formalist* (2021) [175] exemplify this issue.

*Perception Engines* is a series of semi-abstract compositions of percept agglomerations which encapsulate the visual models of various subjects (rabbit, banana, dalmatian, etc.) that score consistent object recognition across different AI vision algorithms. Its pertinent idea of exposing anthropomorphism in AI by emphasizing the pareidolic "universality" of machine vision through a skillful production methodology results in images that prove the concept but are uninteresting. They look like computational deconstructions or rearrangements of hybrid Guston-Warhol-Koons-Murakami-emoji compositions, which document a scientific research project. Printing the images in traditional technique (serigraphy or screen-print) does not make them more convincing, and may even be detrimental if interpreted as a naïve attempt to "legitimize" the work artistically. The poetic identity of *Perception Engines* relies on its clever generative procedure with machine vision algorithms, which is, however, incomprehensible to the most audience.

*Zombie Formalist* critiques digital art's commodification through bland formalism that has been boosted by the crypto art market. In this installation, two AI-powered lightboxes randomly generate images in the style of painters such as Gene David, Barnett Newman, Kenneth Noland, and Karl Benjamin. One box uploads its images to Twitter and records the number of likes and retweets. The other box uses a camera and face detection to generate images when the audience is not looking at it, and to record the attention span for each image when the audience looks at it. Twitter reactions and preferential viewing time are continuously used as variables to train classifiers that establish the difference between "good" (a lot of engagement) and "bad" images (little or no engagement), and to filter generated images favoring "good" ones. By arranging a witty marriage of ML and Komar and Melamid's *People's Choice* (1994–1997) (although Bogart does not acknowledge this referential work [175,176]), *Zombie Formalist* makes a clear case, but mainly for the audience which is already critical of digital art's commodification. For the average audience—which may be unfamiliar with zombie formalism—the project can be counter-effective. If they miss its description or its titular satire, the viewers may take *Zombie Formalist* for what it is functionally: a pair of digital frames that continuously display mildly appealing abstract patterns. As it is often the case with tactical art, the *Zombie Formalist*'s combination of lofty motivation and somewhat ambiguous presentation may diminish its effectiveness or even expedite its recuperation.

This calls for a comparison with Basanta's *All We'd Ever Need . . .* (2018, discussed in Section 2.1.1), which similarly links the notion of "autonomously creative" AI with appropriation strategies. But it couples a complex production setup with a tangible referencing system to provoke mainstream artists, their agents, and collectors instead of "preaching to the choir" by cynical reaffirmation of cultural trends. It effectively critiques the chronic rigidity of intellectual property conventions in general, and particularly the emerging modes of crypto-based art monetization.

3.4.3. Representational Discourse

Critical cogency, viability, and impact of AI art are also affected by the pretentious representational strategies, superficial popular interpretation, and inflated theoretical rhetoric. Following a long-established trend in contemporary art, AI artists are tempted to augment their projects with descriptions that feature elaborate (metaphorical or literal) questions, critical considerations, or theoretical models, but suffer insufficient competence or sincerity. Such strategies can diminish the experience of an artwork by patronizing the audience as pupils rather than independent thinkers capable of appreciating art through their own capacities [6] (pp. 246–247).

This issue is compounded by the (often exaggerated) dependency of AI artworks on external analytical/theoretical discourse, usually provided by academia. However, the intellectual sophistry that theoretical literature invests in identifying the logic or in advocating the plausibility of AI art may also indicate how unintuitive and ethereal its subject has become. Namely, when an artwork requires an explanation or contextualization which is too long, too complex, or too difficult to understand, then it significantly loses its relevance [177].

*3.5. Ethics*

Artists have always faced the challenges of ethical integrity conflicting with professional well-being throughout uneasy coevolution between the open-endedness of artistic proteanism and the ambiguous flux of discourses, criteria, and hierarchies in the artworld and scholarship. AI art reflects the artists' ethical decisions involved in making their works and in building their careers within a context of zeitgeist-relative interferences between the arts, science/technology, cultural tendencies, and socio-political trends [3] (pp. 74–75). Regardless of their modes of involvement with the broader issues of AI ethics, artists are responsible for their own motivations and roles in shaping cultural values and political normalization. They are ethically criticizable when they consciously produce relatively vacuous work or aesthetic "one-liners", when they create derivative projects while dismissing artistic competencies for technical ones, when they downplay the important technical aspects, or when they ignore the socio-political context and implications of the technology. Most artists, authors, and cultural operators prefer to avoid discussing this sensitive territory for the sake of professional survival, which may seem obvious, but in fact draws a higher level of ethical implications. As long as this territory is protected by our hypocrisy and vanity, the cognitive value of art criticism will remain inferior and complacent to diminishing the transformative potentials of the arts.

*3.6. Broader Concerns*

Broader issues that affect contemporary AI art include the uninformed popular discourse, the questionable norms of the art community, the depleting autonomy of academic institutions, and the problematic legal norms for intellectual property and creative labor.

3.6.1. Cultural

The popular writing and media approach toward AI art are still biased toward sensationalistic, superficial, uninformed, or noisy coverage. For example, Klingemann's works, such as *Memories of Passerby I* (2018, a GAN which generates and morphs surreal male and female faces), have been compared to the works of Francis Bacon based on the vague resemblance of mutated portrait elements [25] (p. 248). To anyone with reasonable experience in visual arts, this comparison is untenable and its incompetence could be seen as distasteful. The facial morphs in Klingemann's work are formally disparate, smooth, and lack the paste facture (tactile materiality) which condenses the troubled gestural physicality of Bacon's paintings.

The tendency to conflate AI art practices that represent dilettantism, techno-spectacular fetishism, academic aloofness, deft exploration, and insightful critique into simplistic, often bewildering, or baffling narratives, contaminates both public and theoretical discourse [6] (pp. 241, 245). Additionally, as discussed in Section 2.1, corporate media, some art institutions, authors, and artists misrepresent algorithms as "artists" and uncritically promote shallow or derivative AI art practices. By debasing the artists' cognitive abilities which manifest in less palatable but crucial poetics, they legitimize the regressive, intellectually offensive, and politically dangerous cultural ignorance [178,179] (pp. 7–9).

3.6.2. Professional

Reputation games in the art community are driven by fluid social networks, cliques, coteries, and intrigues, directed by unstable loyalties or affiliations, and shaped by fancy, fashion, and authority appeal. Also, the exceeding topicality and contextual dependence of contemporary art, adds to the (mnemonic) ephemerality of realized artworks [177]. This volatile dynamic tends to virtually reduce merit to a temporal figure of speech while upholding cultural hegemonies, institutional privileges, and profit-driven power games.

Such volatile vocational milieu—combined with inherently high production demands and intrinsic need for endorsement by corporate AI, MCA, or academia—makes AI artists particularly liable to becoming consciously or subconsciously manipulative or cynical, to compromising their creativity, and to softening their critical edge [6] (pp. 252–254). If they strive for integrity, all actors in AI art should be able to recognize these systemically biased and noisy professional value systems, assess them objectively, and correct them.

### 3.6.3. Educational

The globally deteriorating strategic and financial status of education since the 1980s has been causing the erosion of academic autonomy and integrity. It often leads to obedient institutional policies which degrade the educational process from a synergetic coevolution of intensive learning practices through diverse research directives into a routine social service, homogenized by current economic trends and pragmatically focused on vocational training. These policies contribute to the technocentrism in new media art programs (discussed in Section 3.3), which largely remain unconcerned with solving the conflicts between contradictory ideas and opposing world views in the two fields they strive to integrate: modernist in science/technology, and postmodernist in arts and humanities [180] (pp. 92–94). Overall, it seems that systematic effort to recognize, assess, and cultivate students' artistic talent does not match the effort and energy invested in knowledge transfer, administrative work, and program or institution promotion. Furthermore, university-incorporated art schools often enforce inappropriate academic progress evaluation mechanisms cloned from the tech/science community, which affect their faculties' work as educators, researchers, and artists. They have been increasingly tailoring art projects to match the fund application or fund justification strategies and to fit the exhibition-conference-paper pipeline. Such development workflows and presentational formats may be beneficial for some types of art projects, but detrimental to many artworks which are generically incompatible with the procedural or epistemological logic of scientific research. The effect is that in the festival, exhibition, and conference publications, artworks are often mispresented as scientific studies, and vice versa.

### 3.6.4. Proprietary

AI artworks that glamorize narrow or polarized concepts of creativity (human vs. machine, individual vs. collective, etc.) sustain the disputed notions of monolithic authorship rather than advocating for heterogeneous or conjugated actualization of the expressive agency. Intentionally or unintentionally, they reinforce the anthropocentric models of creativity that benefit the problematic culture of proprietary mental labor [22] (p. 135). However, the intellectual work (and ethical values) involved in the design and operation of ML systems are not centralized but distributed. As AI art diversifies, these compound aspects are becoming increasingly evident and addressed more clearly. But narrow definitions of authorship and copyright, and fixed divisions of labor, benefit commercially-driven normative relations of production, distribution, and consumption. The future poetic scope of AI art may be limited by conservative initiatives for imposing legal instruments which would keep the creative decisions under centralized profit-motivated control. The responsibility for tackling these issues lies not only with the artists, but also with scientists, entrepreneurs, cultural agents, and the public.

## 4. Prospects

The AI's technological repertoire and socio-political context both stimulate the artists' creative abilities and reveal their limitations. Contemporary AI art is a burgeoning poetic ecosystem with potentially abundant intellectual and ethical implications, but it is currently relevant primarily for its implicit or explicit reflections of the AI's challenges, shortcomings, and ambiguities.

In *AI, Arts and Design* editorial of Artnodes [4] (p. 3), Burbano and West ask: Does ML creativity in the arts and design represent an evolution of "artistic intelligence", or is it a metamorphosis of creative practice yielding fundamentally distinct forms and modes of authorship? At this point, the critical answer is: neither. Namely, AI artists predominantly approach ML as a toolkit that leverages high computing speed and big data for refined statistical processing of various media; but new media artists have frequently used computationally-driven statistical tools for media processing with similar, often superior, poetic premises and expressive intents before the advent of DL [2,3]. Comparative exemplars in this paper demonstrate this continuation and also show that contemporary AI art largely reiterates various modes of technologically entangled authorship which have already been emphasized by non-AI new media art. These factors substantiate the aesthetic values and overall poetic identity of contemporary AI art, while formal novelties—such as the high degree of mimicry or improved resolution in simulating certain phenomena—are secondary.

Nevertheless, the diversity and criticality of the field have been improving as the initial hype is toning down, and more artists start to use ML. Their investigation of the various aspects and expressions of human intelligence may be their key contribution to AI research and contemporary culture [19]. Most importantly, they can establish insights into all aspects of the AI-influenced world through meaningful relationships with the issues, contingencies, and advances of AI technology, and by considering the expressive logic across the genres of AI art [181]. In order to engage the audience with a lasting impact, AI artists need to balance their motivational sincerity and ideational cogency with procedural skills and maintain a critical outlook on their poetic devices.

*4.1. Competences*

The ethos of maturely calibrated competences deserves cultivation through playfulness, bricolage, technical and conceptual hacking, and imaginative discovery that characterizes other areas of new media art. This ethos comes from the realization that art happens not simply by adding material configurations that no one has witnessed before, but by integrating organized matter into complex human interactions that help us understand the world differently, make us better, or give us a chance to become better. The inherently political nature of AI technology [10] obliges artists not only to exploit but to deconstruct and explore their creative means. They need stronger criteria for poetic thinking, and better multidisciplinary knowledge of historical, theoretical, cultural, and political contexts in which they produce and present their works [182]. They can catalyze their procedural proficiencies by undertaking systematic training in related non-computational art disciplines, so they can appreciate the cognitive and physical demands of creative work in a broader existential sense. By raising the awareness of technocentrism in their practices, AI artists can also promote the necessary changes in STEAM education.

*4.2. Tacticality*

By recognizing and understanding injustices in the notional, relational, technical, political, and other layers of the socio-technological environments in which they live and create, AI artists can overcome cynicism or unconscious resignation that often undermines their critical efforts. The tactical impact can be improved by bolder and more nuanced examination of the cultural and socio-political contexts of AI technology and business, and by deeper probing and problematizing the underlining concepts such as intelligence, creativity, expressive agency, authorship, intellectual labor, ownership, authenticity, accuracy, and bias. The flexibility and mutability of these concepts are inherent to human socio-cultural dynamics and, while technologies such as ML or blockchain challenge them less radically than it is widely presumed [183], they can be used to reconfigure or extend them in interesting and insightful ways.

For tackling the power and sophistication of cultural recuperation in modern infocapitalism (which artists tend to underestimate, overlook, or ignore), stealthy subversiveness and subterfuge seem to be more prudent than didactic overexplanation or overbearing spectacularism. An experiential approach that engages the audience by stimulating their imaginative cognition and critical thinking is often more impactful than surface-based, aestheticized, descriptive, or purely rhetorical artworks that offer a soft or nominal critique of AI-related political power. By demystifying the seemingly radical capabilities of their tools, AI artists can leverage the basic questions and issues of modern AI as critical assets with wide political significance. Empowered by the destabilizing value of humor, a responsible treatment of these assets can build new insights about human nature and provide meaningful posthumanist perspectives [184,185].

*4.3. Creativity*

To address the allurement of exploitatively incentivized creativity (creativity for its own sake) [186], artists should articulate and respect their methodologies as heterogeneous productive frameworks whose processes and outcomes inform the audience by stirring inquisitiveness and critical thinking, by stimulating imagination, and by encouraging progressive action. Within this context, there is an underexplored analogy between the artistically adverse normalization of children's creative idiosyncrasies through socialization, and the artists' conscious or intuitive compliance to cultural trends [3] (pp. 74–75, 77–78). By directing their transgressiveness beyond spite, amusement, or showmanship, artists can turn their wit and versatility into exemplars of meaningful resistance to the socio-political imperatives and existential bleakness [187]. Such exemplars can help us discover new ways to be curious and change our understanding of (being in) the world. By cultivating a dynamic interactive relationship with their progressively sophisticated tools such as ML, artists are in a privileged but also responsible position to push the limits and notions of creativity and in turn inspire the research of computational and technologically augmented creativity.

*4.4. Commitment*

The socio-technical entanglements of AI art with corporate AI, MCA, and academia support the forthcoming art projects, but may also attenuate their criticality and expedite recuperation. A straightforward way for the artists to tackle this precarious relationship is to resist prioritizing their careers over their art, to be open to taking genuine risks, and to pursue systematic support with skepticism toward institutional rationales for art sponsorship. The key requirement of avant-garde art is a deep, constructive dedication to evolving potentially hazardous ideas, and to finding effective ways to share them with the audience. It takes exceptional curiosity, inventiveness, and enthusiasm to do any creative work without anticipating affirmation, compensation, or success in the conventional sense. Artists cannot maintain this costly hierarchy of priorities indefinitely, so the public and the art institutions should rise above their unspoken but unrealistic and ultimately cruel expectance that an artist should steadily deliver significant works.

In a broader prospect, the institutional frameworks of contemporary art, science, technology, and education can provide significant incentives for the unbiased development and representation of AI art, thus enhancing the exploration of AI; but they need thorough reconsideration and reconceptualization in order to be self-critically adaptable for absorbing the knowledge that emerges from various relevant disciplines [6] (p. 255). This requires close cooperation between artists, institutional representatives, and the public in exposing the political hegemonies, and in criticizing the coercive evaluation criteria imposed by the artworld, academia, politics, economy, and the media.

*4.5. A Critical Framework for AI Art*

AI art requires appreciation models for experientially, intellectually, and emotionally competent spectatorship keyed to an artworks' demands [188]. Just as every artistic enterprise must earn the right to claim the audience's most precious assets—time and attention—so the audience needs proper modes of involvement with AI art in order to invest these resources wisely [23].

In a 1995 article titled *Artificial Intelligence Research as Art* [189], Stephen Wilson traced the creative landscape of AI art. He recognized that the artists' relationship with AI reaches beyond technical boundaries towards an investigation into the nature of being human, the nature of intelligence, the limits of machines, and our limits as artefact makers. The critical framework I outlined in this paper informs the investigation of AI art with an inquiry into the accomplishments, shortcomings, and ambiguities across the AI art disciplines, and facilitates comparative insights into their anthropological, socio-political, cultural, and historical aspects. Its principal method is to distinguish the poetic values of AI artworks by acknowledging and contextualizing both their strong and weak points fully respectful of the artists' creative endeavors, but with an awareness that such approach can be misinterpreted as polemical or confrontational. By recognizing the inherently political nature of technology [190] as a default feature of AI, this platform leverages the understanding of processes and infrastructures for the production, presentation, and reception of AI art [191]. That allows it to ask potentially difficult questions about the ethical reasoning behind the artists' creative choices throughout the creation, communication, and cultural stratification of their works. It strives to consider the discourse of AI art objectively and purports to debunk the opaque notions, mystifying verbiage, or dubious claims regardless of the authors' authority. It also requires alertness and self-corrective mechanisms to address the processual challenges of emerging practices, and to combat its own ideological filtering and cognitive flaws such as motivated reasoning. The aesthetic evaluation, as a multidimensional integrative process, is informed by all these considerations [29,192].

This exploratory critique caters to a courageous, inquisitive audience that strives for integrity and inspiration of avant-garde art practices. It respects the intelligence and proteanism that we expect as the artists' key attributes and responsibilities, but it also cultivates an awareness about our tendency to take for granted, overlook, or dismiss the efforts that artists invest in their work. Its aim is to expand the existing critical discourse of AI art with new perspectives for understanding the conceptual and contextual nature of ML as an artistic medium in the age when the arts, together with science and technology, are becoming increasingly responsible for changing ecologies, shaping cultural values, and political normalization. It facilitates rethinking and redefining of the art/science/technology critique, and can be applied to examine the creative attributes of emerging practices in order to assess their cultural significance and socio-political impact. A better understanding of existential conditions, poetic range, and other features of AI art is relevant to both artistic and scientific research in AI, and to its handling by the cultural sector. In a broader prospect, this critical framework can inform the creative actors' reasoning to develop more robust concepts of intelligence, map its perspectives, and make directives for further development and responsible application of AI.

**Funding:** This research received no external funding.

**Institutional Review Board Statement:** Not applicable.

**Informed Consent Statement:** Not applicable.

**Data Availability Statement:** Not applicable.

**Conflicts of Interest:** The author declares no conflict of interest.